\documentstyle[12pt,epsf,epsfig]{article}
\begin{document}
\title{Role of Bose enhancement in
photoassociation\footnote{Paper submitted to
the special issue of J. Mod. Opt. on ``Seminar on Fundamentals
of Quantum Optics V'', K\"{u}htai, Austria, January 16-21, 2000;
ed. F. Ehlotzky.}}
\author{Matt Mackie and Juha Javanainen\\ Department of
Physics, University of Connecticut\\ Storrs, CT 06269-3046, USA}
\maketitle
\abstract{We discuss the role of Bose enhancement of the dipole
matrix element in photoassociation, using stimulated Raman
adiabatic passage as an example. In a
nondegenerate gas the time scale for coherent optical
transients tends to infinity in the thermodynamic limit, whereas
Bose enhancement keeps this time scale finite in a condensate.
Coherent transients are therefore absent in photoassociation of
a thermal nondegenerate gas, but are feasible if the gas is a
condensate.}
\baselineskip 24pt

\section{Introduction} The theoretical realization is now
emerging that it may be possible to effect analogs of coherent
optical transients, such as Rabi flopping~\cite{DRUM98,JJMM99},
rapid adiabatic passage~\cite{JJMM99}, and stimulated Raman
adiabatic passage (STIRAP)~\cite{BSTIRAP}, in photoassociation
of a Bose-Einstein condensate (BEC). In contrast, the
feasibility of transients in PA of a thermal nondegenerate gas
has been controversial. There are predictions of STIRAP in
ordinary gases~\cite{VARDI97}, whereas our results suggest
otherwise~\cite{JJMM98}.

The purpose of the present paper is to clarify the status of
coherent optical transients in both nondegenerate and
degenerate thermal gases. Our key concept is Bose enhancement.
Suppose we have a large number of atoms $N$ in one quantum
state, as in a BEC~\cite{BECREV}. Due to Bose-Einstein
statistics, the transition matrix element referring to the BEC
will pick up a multiplicative factor
$\sqrt{N}$ as compared to the matrix element for a single atom.
It turns out that Bose enhancement will make a
difference. In the thermodynamic limit, when the
volume and particle number of a gas tend to infinity while the
density remains constant, coherent optical transients will
vanish in photoassociation of a nondegenerate gas, but
Bose enhancement will facilitate such transients in a
condensate. In the rest of the present paper we outline the
details of this argument using STIRAP as our explicit example.

\section{Ordinary STIRAP} 
\begin{figure}
\centering
\epsfig{file=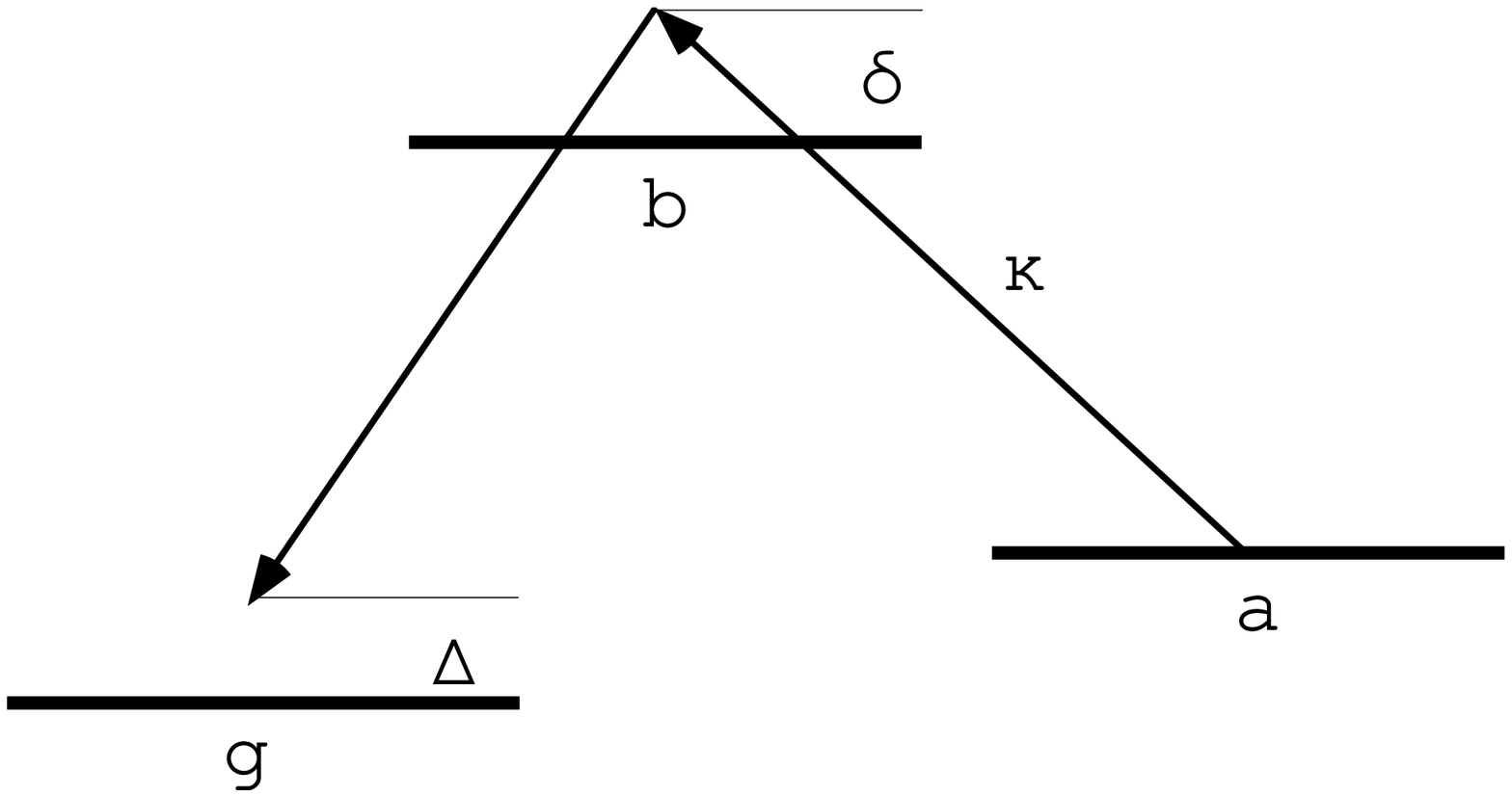,width=8cm,height=13cm}
\caption{Three-level $\Lambda$ scheme with states $a$, $b$ and
$g$ and laser couplings with Rabi frequencies $\kappa$ and
$\Omega$. The definitions of the two-photon and intermediate
detunings $\Delta$ and
$\delta$ are also given schematically. For STIRAP, one would
have
$\Delta=0$. The notation also applies to the
Hamiltonian~(\protect\ref{H}).}
\label{STIRAP}
\end{figure}

As a prelude to our development we
briefly reiterate the salient features of STIRAP in an ordinary
three-level $\Lambda$ system~\cite{STIRAP}, as in
Fig.~\ref{STIRAP}. Two laser fields are tuned to exact
two-photon resonance between the two lower states $a$
and $g$ of the
$\Lambda$ system. First the laser intensities are arranged in
such a way that the coupling between states $g$ and $b$ is much
stronger than the coupling between $a$ and $b$. One of the
dressed states of the three-level system then
coincides with the bare state
$a$. When the laser intensities are adjusted in such a way that
the coupling becomes much stronger in the transition from
$a$ to $b$, the dressed state that initially coincided with the
bare state $a$ turns instead into the bare state $g$.

Suppose now that, before the adjustment of laser
intensities, the system started in the bare state
$a$. It then also started in the dressed state whose fate we
follow. If the lasers were adjusted slowly enough, adiabaticity
guarantees that the system stays in the dressed
state all along. In the end, the system therefore winds up in
the bare state $g$. Moreover, a detailed analysis shows that, in
the ideal adiabatic limit, there is never any population at all
in the intermediate state $b$.

In short, the intensities of the two light pulses are switched
in such a way that initially the coupling between the unoccupied
states is much stronger than the coupling to the occupied
state, and the same situation prevails when the coupling
strengths have reversed. This is called \underline{counter-intuitive}
pulse order. The result is that the system is transferred from the
initial state
$a$ to the final state $g$. Besides, ideally, the system never
visits the intermediate state $b$, which is a major virtue if
$b$ is plagued by dissipation.

\section{Quasicontinuum photoassociation} In a process
of photoassociation two atoms and a photon combine to make the
corresponding diatomic molecule. As the two atoms may be
considered to be a dissociated state of the molecule,
photoassociation is really about free-bound transitions. There
are internal atomic and molecular states involved in photoassociation
as well, but we assume that, by properly choosing the laser
frequency, one has selected a transition from the initial state
of the atoms to a unique rovibrational state of the molecule.

Quantum optics normally deals with bound-bound transitions
responsible for characteristic spectral lines, and with
bound-free transition that manifest themselves in decay
processes such as spontaneous emission. From this
angle, photoassociation is irreversible decay in reverse, and
may seem to violate the second law of thermodynamics.
Nonetheless, photoassociation spectroscopy is alive and well,
and is the source of the most accurate molecular structure data
available at this time~\cite{PAREV}.

To model free-bound transitions, especially in a thermal
sample, it is desirable to start with two atoms whose relative
motion is in an eigenstate of energy. Unfortunately, as energy
eigenstates of the relative motion for unconfined atoms are not
normalizable, there is no mathematically sound way to write down
such a quantum state. The cure is straight from textbooks of
quantum mechanics: Assume that the relative motion is
restricted to a quantization volume $V$, and at the end of the
calculations take the limit $V\rightarrow\infty$.

This is the stratagem of our quasicontinuum (QC)
approach~\cite{JJMM98,MMJJ99}. The dissociation continuum is
broken up into discrete states, a QC, whereupon problems with
the normalization of the states vanish. In an unexpected
windfall, the mathematics also turns out to work out in such a
way that the experience about few-level systems gained over
decades of quantum optics and laser spectroscopy is directly
transferable to understanding photoassociation.

As it comes to the theme of the present paper, the
observation of most immediate relevance is that the dipole
matrix element between any free (two-atom) state and the bound
(rovibrational molecular) state scales with the quantization
volume as $\,d\sim V^{-1/2}$. There is no mystery to this. A
unit-normalized dissociated state fills the entire volume
$V$, so its square is proportional to $1/V$ and the
normalization constant is $\propto V^{-1/2}$. On the other
hand, the normalization constant of the bound state is
independent of $V$. Eventually the
$V^{-1/2}$ makes its way to the free-bound dipole matrix
element.

The immediate consequence is that in the limit of an infinitely
large quantization volume, the dipole matrix element between
the bound state and any single QC state vanishes. Accordingly,
when the volume is increased and Bose enhancement is absent,
the photoassociation coupling can always be treated with perturbation
theory~\cite{MMJJ99}.

\section{Nondegenerate thermal gas}
\subsection{Theoretical method}

The main reason why we have insisted on eigenstates of energy
is the basic random-phase postulate of statistical mechanics,
which states that the thermal density operator is diagonal in
eigenstates of energy. It is therefore always permissible to do
whatever analysis one is aiming at by first assuming that the
system starts in a given eigenstate of energy $m$, and at the
end of the calculations averaging the results over the thermal
probability distribution of the states $m$.

When discussing a nondegenerate thermal gas, we make another
assumption as well. We analyze photoassociation for
just two atoms, and calculate the free-bound transition rate,
$R$. However, a typical experiment involves
$N\gg2$ atoms. When thinking of
photoassociation of any given `probe' atom, we add the
transition rates due to all colliders, so that the photoassociation rate per
atom becomes $NR$.

In the limit of infinite quantization volume, the rate of photoassociation
for two atoms vanishes with the coupling matrix
element as
$R\propto d^2\propto 1/V$. We of course expect as much since
two atoms cease to collide in an infinite volume. Nonetheless,
for
$N$ atoms the photoassociation rate per atom is  $ NR\propto N/V =
\varrho$, proportional to the density of the gas. In the 
\underline{thermodynamic limit} when both $N$ and $V$ tend to infinity
in such a way that $\varrho$ remains constant, the photoassociation rate per
atom has a finite limit proportional to density. The result is
reasonable, and our QC approach~\cite{JJMM98,MMJJ99} in fact
exactly reproduces the free-bound transition rate
obtained from collision theory~\cite{PAREV}.

\subsection{The demise of STIRAP}

\begin{figure}
\centering
\epsfig{file=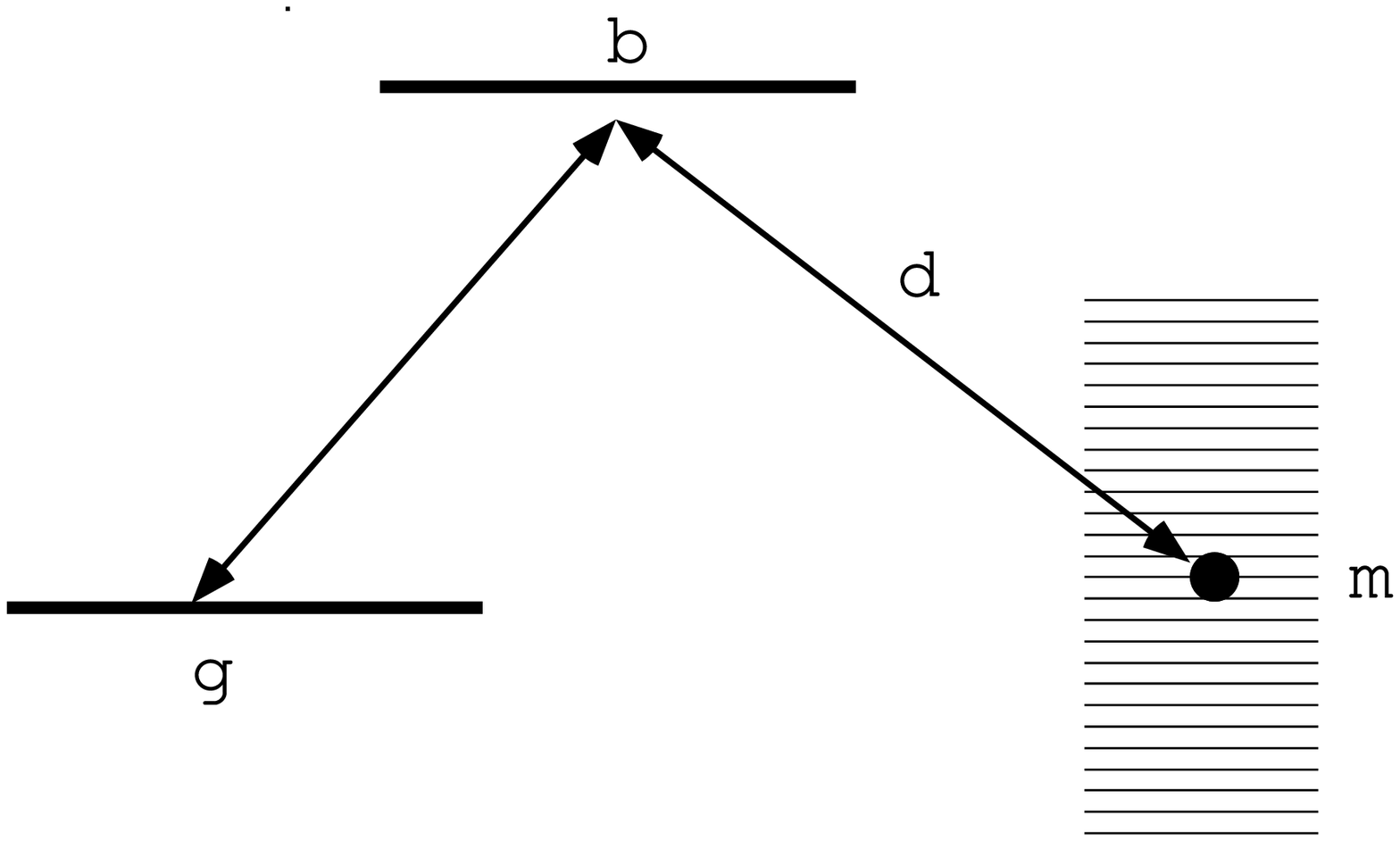,width=8cm,height=10cm}
\caption{Scheme for STIRAP starting from the quasicontinuum
state $m$. The free-bound coupling is represented here by the
dipole matrix element, $d$.}
\label{QCSTIRAP}
\end{figure}

In our model, and in current practice~\cite{PAREV}, the primary
photoassociated molecular state is reached by absorption of a
photon. This invariably means that the molecule is subject to
spontaneous emission, and decays away. Moreover, the reverse of
PA, photodissociation, tends to break molecules back into
atoms. It is then a natural idea to add a second laser field
tuned between the initially photoassociated state and another
(more) stable molecular state, and to attempt to utilize STIRAP
to avoid losses from the primary photoassociated
state~\cite{VARDI97}. We sketch such a scheme in
Fig.~\ref{QCSTIRAP}.

Unfortunately, this idea hits a roadblock. According to the
random-phase postulate, for a thermal sample we may assume that
the system starts in a given QC state
$m$, as denoted in Fig.~\ref{QCSTIRAP}. But the free-bound
matrix element $d$ scales with the quantization volume $V$ as
$d\propto V^{-1/2}$, and tends to zero with
$V\rightarrow\infty$. On the contrary, the bound-bound matrix
element is volume independent. The free-bound coupling is
therefore \underline{always} small compared to the coupling for
bound-bound transitions, and it is impossible to effect the
counter-intuitive reversal of the coupling strengths needed for
STIRAP~\cite{JJMM98}. Of course, the final average over a
distribution of initial states
$m$ is not expected to create STIRAP either.

We are not arguing that there would be \underline{no} two-color
photoassociation; see Ref.~\cite{MMJJ99}. However, in
the absence of any evidence to the contrary, we do not believe
that the advantages of counter-intuitive pulse order and STIRAP,
protection from decay of the intermediate state and the ensuing
improvement in transfer efficiency, will materialize in
free-bound-bound photoassociation of a thermal nondegenerate gas.

\section{Bose-Einstein condensate}
\subsection{Theoretical method}

Calculating the free-bound transition rate for one pair of
atoms and then multiplying by the number of available colliders
is a process which implicitly assumes that we can distinguish
between the atoms. Such an approach is fundamentally flawed in
the case of a BEC. Instead, we have adopted a phenomenological
second-quantized Hamiltonian for
photoassociation~\cite{DRUM98,JJMM99}.

Our basic premises are that one may treat atoms and molecules as
bosons in their own right, and that photoassociation conserves
momentum. Given the momentum representation for atoms and
molecules, annihilation operators
$a_{\bf k}$ and
$b_{\bf k}$, the part of the Hamiltonian responsible for
photoassociation reads,
\begin{equation} H = \ldots - \hbox{$1\over 2$}\sum_{{\bf k},
{\bf p}, {\bf q}} {\bf d}({\bf k}-{\bf p})\cdot {\bf E}_{\bf
q}\,b^\dagger_{{\bf k}+{\bf p}+{\bf q}}a_{\bf k}a_{\bf p}+
\ldots\,.\label{H}
\end{equation} Here ${\bf E}_{\bf q}$ is the Fourier
component
${\bf q}/\hbar$ of the positive frequency part of the electric
field driving photoassociation. By translational symmetry, the
dipole matrix element may depend only on the difference of the
momenta of the atoms, ${\bf k}-{\bf p}\,$, and in the dipole
approximation it cannot depend on the photon momentum $\bf
q$. The term written down is simply a sum of processes in which
two atoms with momenta $\bf k$, $\bf p$ and a photon with
momentum $\bf q$ are combined into a molecule with momentum
${\bf k}+ {\bf p}+{\bf q}$.

It remains to determine the values of the dipole matrix
elements $\bf d$ in the Hamiltonian. We do this by demanding
that for the nondegenerate thermal gas the results from the
Hamiltonian~(\ref{H}) be the same as we obtain from our QC
approach. In the process a number of subtleties come up having
to do, e.g. with Bose-Einstein statistics and the Wigner
threshold law for photodissociation~\cite{JJMM99,MK00}. The
bottom line, though, is that we know how to deduce the matrix
elements from considerations such as the standard
molecular-structure calculations, or
measurements of the photodissociation rate.

Consider now an ideal zero-momentum condensate photoassociated
by a plane wave of light, where all photons have momentum
$\bf q$. The molecules made by photon absorption
all have the momentum $\bf q$. The converse is not trivially
true. By momentum conservation alone, the induced emission of a
molecule with momentum $\bf q$ need not return two atoms
into the condensate; but, in a process we call rogue
photodissociation~\cite{MK00}, the two atoms may emerge with any
opposite nonzero momenta. However, the
photodissociation processes that return the atoms back to the
condensate are favored by Bose enhancement, and rogue
photodissociation is further suppressed by energy
conservation. Thus we adopt a two-mode model, only taking into
account atoms with zero momentum and molecules with momentum
$\bf q$. The corresponding creation and annihilation
operators are denoted by $a$ and~$b$.

\subsection{The return of STIRAP}

Turning now towards two-color photoassociation of a degenerate
gas~\cite{HEI00b}, we reconsider the possibility of
STIRAP~\cite{BSTIRAP}. It is thus assumed that a further laser beam
couples the primarily photoassociated molecule to another bound
molecular state, whose annihilation operator is denoted by $g$. The
three-mode Hamiltonian reads
\begin{equation} {H\over\hbar} = -\Delta g^\dagger g - \delta
b^\dagger b - \hbox{$1\over 2$} \kappa (b^\dagger aa +b
a^\dagger a^\dagger) - \hbox{$1\over 2$} \Omega (b^\dagger g +b
g^\dagger)\,.
\label{HAM}
\end{equation}
Here $\Delta$ and $\delta$ are the two-photon
and intermediate detunings, including the proper photon recoil
energies, the free-bound QC Rabi frequency is
\begin{equation}
\kappa = {{\bf d}\cdot {\bf E}\over 2\hbar}
\end{equation}
and $\Omega$ is the bound-bound Rabi frequency.

Given the Hamiltonian~(\ref{HAM}), the Heisenberg equations of
motion for the boson operators read
\begin{eqnarray}
\dot{a} &=& i\kappa\,a^\dagger b\,,\\
\dot{b} &=& i\delta\, b  +\hbox{$1\over2$}i(\kappa\, aa
+\Omega\, g)\,,\\
\dot{g} &=& i\Delta\, g+\hbox{$1\over2$}i\Omega\, b\,.
\end{eqnarray}

Suppose now that, if all molecules were
dissociated to atoms, there were $N$ atoms. The boson operators
in the system are then of the order
$\sqrt{N}$. We thus introduce the rescaled boson operators
$\alpha = a/\sqrt{N}$,~\ldots. These operators are  of the
order of unity, and may roughly be interpreted as the
second-quantized counterparts of the probability amplitudes
that an atom is in the atomic condensate ($\alpha$) or in one
of the two molecular condensates ($\beta$, $\gamma$). The
scaled operators obey the equation of motion
\begin{eqnarray}
\dot{\alpha} &=& i\chi\, \alpha^\dagger\beta\,,\label{EQA}\\
\dot{\beta} &=& i\delta\, \beta + \hbox{$1\over2$}i(\chi\,
\alpha\alpha +\Omega\, \gamma)\,,\\
\dot{\gamma} &=& i\Delta\,\gamma +\hbox{$1\over2$}i\Omega\,
\beta\,.\label{EQC}
\end{eqnarray}

The key point of our argument emerges from an inspection of the
new Rabi frequency after scaling,
\begin{equation}
\chi = {\sqrt{N}{\bf d}\cdot{\bf E}\over2\hbar}\,.
\end{equation} The $\sqrt{N}$ is nothing but Bose enhancement
in the present context. In the thermodynamic limit ${\bf
d}\propto\sqrt{1/V}$, so that
$\chi
\propto \sqrt{N/V} = \sqrt{\rho}$ remains finite. More
precisely, suppose that the same laser field with amplitude
$\bf E$ were tuned in such a way that the the photodissociation
of bound molecules would produce two atoms with the reduced
mass $\mu$ and relative velocity $v$ at the rate $\Gamma(v)$,
then we have~\cite{JJMM99,MK00}
\begin{equation}
\chi =
\lim_{v\rightarrow0}\,\sqrt{2\pi\hbar^2\Gamma(v)\varrho\over\mu^2v}\,.
\label{KAPPA}
\end{equation} The limit is finite and nonzero by virtue of the
Wigner threshold law.

The couplings $\chi$ and $\Omega$ in
Eqs.~(\ref{EQA})-(\ref{EQC}) do not depend on the quantization
volume anymore; the dependence on
$V$ is replaced by a dependence on $\varrho$, the density of
atoms if all molecules were to dissociate. In a BEC there is no
longer any intrinsic restriction on the relative size of the
couplings, and photoassociative STIRAP is feasible even in the
thermodynamic limit.

We have constructed explicit examples of STIRAP by solving Eqs.
(\ref{EQA})-(\ref{EQC}) in a semiclassical or
mean-field~\cite{BECREV} approximation, treating
$\alpha$,
$\beta$ and
$\gamma$ as
$c$-numbers instead of quantum operators~\cite{BSTIRAP}. Unlike
the equations for probability amplitudes for the ordinary
$\Lambda$ system, these equations are nonlinear. Nonetheless,
the basic character of STIRAP remains intact. In particular, for
completely adiabatic switching of pulse strengths (for which
Ref.~\cite{BSTIRAP} gives quantitative criteria), there is never
any probability in the primary photoassociated state.
Spontaneous-emission losses and rogue photodissociation are
then shut off.

We cannot think of any matter of principle that could go
catastrophically wrong with STIRAP. In practical experiments the
atoms and possibly also the molecules are trapped and do not
form infinite homogeneous condensates. Trapped particles come
with a time scale that is of the order of tens of milliseconds
in the magnetic case. If the STIRAP pulses are faster,
trapping should not matter a whole lot. We have also ignored
collisions between atoms, between molecules, and between atoms
and molecules. But again, the corresponding time scales could
easily be much longer than the time scale of the laser pulses,
so that collisions  are also negligible. The worst practical
enemy of STIRAP might be light shifts of the
two-photon resonance, which come about as a result of virtual
transitions accompanying rogue photodissociation~\cite{MMJJ99,MK00}.
However, in principle one can always compensate for light
shifts by chirping the laser pulses.

\section{Ruminations}

Our renunciation of photoassociative STIRAP in a thermal
nondegenerate gas most likely applies to other coherent optical
transients as well~\cite{MMJJ99}. In the limit
$V\rightarrow\infty$ the free-bound Rabi frequency
$\kappa\propto1/\sqrt{V}$ tends to zero, and the corresponding
time scale for the transient $1/\kappa$ tends to infinity. At
least in a large enough sample, something else is likely to
happen before a coherent transient can run its course.

Similarly, coherent transients should work in photoassociation
of a BEC. Because of Bose enhancement, the coupling matrix
element picks up a the factor
$\sqrt{N}$, giving the effective Rabi frequency
$\chi=\sqrt{N}\kappa\propto\sqrt{N/V}=\sqrt{\varrho}$. In the
thermodynamic limit $\chi$ remains finite, and the time scale of
transients actually scales with atom density as $\rho^{-1/2}$.
Such transients have already been discussed in the
photoassociation literature~\cite{DRUM98,JJMM99}, and
equivalent mathematics is involved in a wide variety of topics
ranging from second-harmonic generation~\cite{SHG} to the
Feshbach resonance~\cite{EDDY99}.

One final point concludes our discussion of transients in
photoassociation. Bose enhancement oftentimes makes little
practical difference even when it is formally present. For
instance, suppose that we were to calculate a transition rate
for a condensate under a normal bilinear coupling of the form
$b^\dagger a$. Bose enhancement gives the factor $\sqrt{N}$ in
the matrix element, and the transition rate for the condensate
in terms of the one-atom rate $R$ reads $NR$. However, in the
second quantized formulation all $N$ condensate atoms are
treated at once, so that the transition rate \underline{per atom}
is still $R$. On the other hand, the coupling for
photoassociation,
$b^\dagger a a$, is trilinear. The ensuing Bose enhancement
$\propto(\sqrt{N})^2$ is thus stronger than for a bilinear
coupling; in fact, just strong enough to offset the decrease of
the coupling matrix element in the thermodynamic limit. As a
result of the confluence of BEC and cubic coupling, a condensate
responds to photoassociation as a unit on a time scale that
depends on density, but oddly enough not directly on atom
number or size of the condensate.

\section{Acknowledgements}
One of us (JJ) specifically thanks for Prof. Fritz Ehlotky for
the opportunity to participate in the Obergurgl/K\"{u}htai
meetings already for the fifth time --- and hopefully
running\ldots. This work is supported in part by NSF, Grant No.
PHY-9801888, and by NASA, Grant No.\ NAG8-1428.


\newpage



\begin{thebibliography}{99}

\bibitem{DRUM98} Drummond, P. D., Kheruntsyan, K. V., and He, H.,
1998, {\it Phys. Rev. Lett.}, {\bf 81}, 3055.

\bibitem{JJMM99} Javanainen, J., and Mackie, M., 1999,
{\it Phys. Rev. A}, {\bf 59}, R3186.

\bibitem{BSTIRAP} Mackie, M., Kowalski, R., and Javanainen, J., 2000,
{\it Phys. Rev. Lett.}, {\bf 84}, 3803;
http://arXive.org/abs/physics/9909060.

\bibitem{VARDI97} Vardi, A., Abrashkevich, D., Frishman, E., and
Shapiro, M., 1997, {\it J. Chem. Phys.}, {\bf 107}, 6166.

\bibitem{JJMM98} Javanainen, J., and Mackie, M., {\it Phys. Rev. A},
1998, {\bf 58}, R789.

\bibitem{BECREV} Parkins, A. S., and Walls, D. F., 1998,
{\it Phys. Rep.}, {\bf 303}, 1.

\bibitem{STIRAP}  Bergman, K., Theuer, H., and Shore, B. W., 1998,
{\it Rev. Mod. Phys.}, {\bf 70}, 1003.

\bibitem{PAREV} Weiner, J., Bagnato, V. S., Zilio, S. C., and 
Julienne, P. S., {\it Rev. Mod.} {\it Phys.},~{\bf 71},~1.

\bibitem{MMJJ99} Javanainen, J., and Mackie, M., {\it Phys. Rev. A},
1999, {\bf 60}, 3174.

\bibitem{MK00} Ko\u{s}trun, M., Mackie, M., C\^{o}t\'{e}, R., and
Javanainen, J., http://arXive.org/abs/ physics/0006066.

\bibitem{HEI00b} Wynar, R., Freeland, R. S., Han, D. J., Ryu, C.,
and Heinzen, D. J., 2000, {\it Science}, {\bf 287}, 1016.

\bibitem{SHG} Walls, D. F., and Barakat, R., {\it Phys. Rev.A}, 1970,
{\bf 1}, 446; Walls, \mbox{D. F.}, {\it Phys. Lett.}, 1970,
{\bf 32A}, 476; Walls, D. F., and Tindle, C. T., {\it Lett. Nuovo
Cimento}, 1971, {\bf 2} 915; Walls, D. F., and Tindle, C. T., {\it
J. Phys. A}, 1972, {\bf 5}, 534.

\bibitem{EDDY99} Timmermans, E., Tommasini, P., C\^{o}t\'{e}, R.,
Hussein, M., and Kerman, A., {\it Phys. Rev. Lett.}, 1999, {\bf 83},
2691.

\end{thebibliography}
\end{document}